\documentclass[pre,aps,preprint,showpacs,preprintnumbers,amsmath,amssymb,secnumroman,eqsecnum]{revtex4}

\usepackage{graphicx}
\usepackage{dcolumn}
\usepackage{bm}

\newcommand{\beq}{\begin{equation}}
\newcommand{\eeq}{\end{equation}}
\newcommand{\beqa}{\begin{eqnarray}}
\newcommand{\eeqa}{\end{eqnarray}}
\newcommand{\vc}[1]{\mbox{\boldmath $#1$}}

\newcommand{\vol}[1]{{\bf #1}}

\newcommand{\du}[1]{{\bf\sf #1}}

\begin{document}


\title{Effect of fluid inertia on the motion of a collinear swimmer}

\author{B. U. Felderhof}

 \email{ufelder@physik.rwth-aachen.de}
\affiliation{Institut f\"ur Theorie der Statistischen Physik\\ RWTH Aachen University\\
Templergraben 55\\52056 Aachen\\ Germany\\
}%



\date{\today}

\begin{abstract}
The swimming of a two-sphere system and of a three-sphere chain in an incompressible viscous fluid is studied on the basis of simplified equations of motion which take account of both Stokes friction and added mass effects. The analysis is based on an explicit expression for the asymptotic periodic swimming velocity and a corresponding evaluation of the mean rate of dissipation. The mean swimming velocity of the two-sphere system is found to be non-vanishing provided that the two spheres are not identical. The swimming of a comparable chain of three identical spheres is much more efficient.
\end{abstract}

\pacs{47.15.G-, 47.63.mf, 47.63.Gd, 87.17.Jj}
\maketitle
\section{\label{I}Introduction}

In earlier work we studied the effect of inertia on laminar swimming and flying of an assembly of rigid spheres in an incompressible viscous fluid \cite{1}. The spheres were assumed to interact directly via central forces, and hydrodynamic interactions were calculated from low Reynolds number hydrodynamics \cite{2} and the theory of potential flow \cite{3}. It turned out that in a wide range of dimensionless viscosity the swimming is similar to that found in the Stokes regime where friction dominates \cite{4}. In the following we extend our study on the basis of an explicit expression for the asymptotic periodic swimming velocity.

The scallop theorem valid in the Stokes regime \cite{5} shows that a body consisting of two spheres, immersed in a viscous fluid and in slow relative oscillatory motion along the connecting axis, does not swim. It was recently found by Klotsa et al. in experiment and computer simulation \cite{6} that such a body does swim for fast relative motion, provided the spheres are not identical. This occurs in the inertial regime, where mass effects dominate.

For the model with point hydrodynamic interactions we find a non-vanishing mean swimming velocity for the two-sphere system, even though the mean impetus vanishes. However, the transition seen in experiment \cite{6} is beyond the scope of the present theory. Apparently, for a sufficiently large value of the streaming Reynolds number a new mechanism sets in with a transcritical bifurcation type instability \cite{7}, leading to effective swimming in the inertial regime.

It is of interest to study the effects of inertia on swimming outside the Stokes regime, but before the transition. The Stokes regime corresponds to the swimming of microorganisms. Many small organisms occurring in nature will fit the intermediate regime considered here.

The two-sphere system with collinear motion is of theoretical interest, but it does not swim at all in the Stokes regime and is fairly inefficient when inertial effects are taken into account. We investigate also the effects of inertia on the swimming of a collinear three-sphere chain. This extends earlier work in the Stokes regime \cite{8},\cite{9}. The analysis suggests that the theory can be applied also to more complicated sphere models with realistic hydrodynamic interactions.

In Sec. II we derive the general expression for the asymptotic periodic swimming velocity in the discrete sphere model. The expression demonstrates that inertia leads to an after-effect in the relation between swimming velocity and impetus \cite{1}, whereas in the Stokes regime the  relation is instantaneous. As a consequence, the swimming velocity depends significantly on the mass densities of spheres and fluid.

In the intermediate regime, where both friction and inertia are relevant, the swimming velocity depends in a remarkably complicated fashion on the amplitude and period of the stroke. On the other hand, the present work shows that the numerical dependence on parameters and amplitude is fairly simple for both the two-sphere swimmer and the three-sphere chain.

\section{\label{II}$N$-sphere swimmer}

We consider $N$ spheres of radii $(a_1,...,a_N)$  aligned along the $x$ axis of a Cartesian system of coordinates and immersed in a viscous incompressible fluid of shear viscosity $\eta$ and mass density $\rho$. The spheres are assumed to be uniform with mass densities $(\rho_{p1},...,\rho_{pN})$. The fluid flow velocity $\vc{v}(\vc{r},t)$ and pressure $p(\vc{r},t)$ are assumed to satisfy the Navier-Stokes equations and the flow velocity is assumed to satisfy the no-slip boundary condition at the surface of each sphere. The spheres interact via central direct interaction forces which depend only on the relative distances $|x_j-x_k|$ of sphere centers, and they are acted upon by forces $(\vc{E}_1(t),...,\vc{E}_N(t))$ oscillating in time with period $\mathcal{T}=2\pi/\omega$ and directed along the $x$ axis, so that $\vc{E}_j(t)=E_j(t)\vc{e}_x$. Moreover, we require that at any time the applied forces sum to zero, so that the total force acting on the assembly vanishes. If the spheres are initially at rest and located on the $x$ axis, then by symmetry they will remain on the axis, so that it suffices to consider the $x$ coordinates of the centers. We are interested in the asymptotic periodic motion of the center
\begin{equation}
\label{2.1}X(t)=\frac{1}{N}\sum^N_{j=1}x_j(t).
\end{equation}
The swimming velocity is  defined as $U_{sw}(t)=dX/dt$ at long times. Its average over a period at long times defines the mean swimming velocity $\overline{U}_{sw}$.

We assume that at any time the hydrodynamic interactions between the spheres can be described approximately by a $(N-1)\times(N-1)$ friction matrix $\vc{\zeta}$, which can be evaluated from the steady state Stokes equations, and by a $(N-1)\times(N-1)$ mass matrix $\du{m}$, which can be evaluated from potential flow theory. The two matrices depend only on the relative coordinates. In vector notation with $\du{R}=(x_1,...,x_N)$ and $\du{U}=(\dot{x}_1,...,\dot{x}_N)$ the sphere momenta $\du{p}=(p_1,...,p_N)$ are given by \cite{1}
\begin{equation}
\label{2.2}\du{p}=\du{m}\cdot\du{U},\qquad \du{U}=\du{w}\cdot\du{p},
\end{equation}
where $\du{w}=\du{m}^{-1}$ is the inverse mass matrix. We assume that to a good approximation the motion of the spheres is described by the equations of motion \cite{1}
\begin{equation}
\label{2.3}\frac{d\du{R}}{dt}=\du{U},\qquad\frac{d\du{p}}{dt}=-\frac{\partial\mathcal{K}}{\partial\du{R}}-\vc{\zeta}\cdot\du{U}-\frac{\partial V_{int}}{\partial\du{R}}+\du{E},
\end{equation}
where $\mathcal{K}$ is given by $\mathcal{K}=\frac{1}{2}\du{p}\cdot\du{w}\cdot\du{p}$ and $V_{int}$ is the potential of direct interaction forces. The partial derivative $\partial/\partial\du{R}$ is taken at constant momenta $\du{p}$. The applied forces are summarized in $\du{E}=(E_1,...,E_N)$. It follows from Eq. (2.3) that the center velocity $U(t)=dX/dt$ satisfies the equation of motion \cite{1}
\begin{equation}
\label{2.4}\frac{d}{dt}\;(MU)+ZU=\mathcal{I}
\end{equation}
with time-dependent mass $M(t)$ and friction coefficient $Z(t)$ given by
 \begin{equation}
\label{2.5}M=\du{u}\cdot\du{m}\cdot\du{u},\qquad Z=\du{u}\cdot\vc{\zeta}\cdot\du{u},\qquad\du{u}=(1,...,1),
\end{equation}
and impetus $\mathcal{I}(t)$ given by
\begin{equation}
\label{2.6}\mathcal{I}(t)=-\frac{d}{dt}(\du{u}\cdot\du{m}\cdot\dot{\du{d}})-\du{u}\cdot\vc{\zeta}\cdot\dot{\du{d}},
\end{equation}
where $\dot{\du{d}}$ is the time-derivative of the displacement vector $\du{d}(t)$, which is defined such that $\du{u}\cdot\du{d}(t)=0$. The vector will be specified below.

From Eq. (2.4) we find for the asymptotic periodic swimming velocity
\begin{equation}
\label{2.7}U_{sw}(t)=\frac{1}{M(t)}\int^t_{-\infty}\exp\big[-\int^t_{t'}\frac{Z(t'')}{M(t'')}\;dt''\big]\mathcal{I}(t')\;dt'.
\end{equation}
The impetus $\mathcal{I}(t)$ acts as a driving force and determines the swimming velocity with a time-lag. In the Stokes limit inertia is neglected, and then the swimming velocity is given simply by
\begin{equation}
\label{2.8}U_{sw}(t)=\frac{-1}{Z(t)}\;\du{u}\cdot\vc{\zeta}\cdot\dot{\du{d}},\qquad (\mathrm{Stokes}),
\end{equation}
with no time-lag. The time-lag in Eq. (2.7) becomes infinite in the limit of small viscosity.

 It is useful to measure the influence of viscosity in terms of the dimensionless scale number $s$  given by \cite{10}
\begin{equation}
\label{2.9}s^2=\frac{a^2\omega\rho}{2\eta},
\end{equation}
where $a$ is a typical radius. The Stokes limit corresponds to $s\rightarrow 0$. The scale number is related to the Roshko number by $Ro=4s^2/\pi$, if in the latter we use the diameter $2a$ as the characteristic length.

The drag force exerted on the spheres by the fluid is given by
\begin{equation}
\label{2.10}D(t)=-\du{u}\cdot\vc{\zeta}\cdot\du{U}=-ZU_{sw}-\du{u}\cdot\vc{\zeta}\cdot\dot{\du{d}}.
\end{equation}
In periodic swimming the time-average of the drag over a period vanishes, as follows from Eq. (2.4) and the above definition. The mean impetus $\overline{\mathcal{I}}$, where the overhead bar indicates the average over a period, in general does not vanish.

The time-dependent rate of dissipation is given by
\begin{equation}
\label{2.11}\mathcal{D}=\du{U}\cdot\vc{\zeta}\cdot\du{U}.
\end{equation}
The power used for a stroke in periodic swimming is
\begin{equation}
\label{2.12}P=\overline{\mathcal{D}}.
\end{equation}
It is of interest to study the mean swimming velocity $\overline{U_{sw}}$ for given power $P$ as a function of the parameters.

We set ourselves the goal of calculating the asymptotic periodic swimming velocity $U_{sw}(t)$ for given periodic relative motion $\du{r}(t)$. This is a kinematic point of view. Since the friction matrix $\vc{\zeta}$ and the mass matrix $\du{m}$ depend only on relative coordinates the time-dependent total friction coefficient $Z(t)$ and the total mass $M(t)$ can be calculated from Eq. (2.5) once $\du{r}(t)$ is specified. The time-dependent impetus $\mathcal{I}(t)$ can be evaluated from Eq. (2.6). These three quantities are sufficient to calculate the swimming velocity $U_{sw}(t)$ from Eq. (2.7). The applied periodic forces $\du{E}(t)$ necessary to achieve the motion may be evaluated subsequently from the equations of motion (2.3). For small amplitude motion the relative motion $\du{r}(t)$ which leads to maximal speed for given power can be determined from an eigenvalue problem.

\section{\label{III}Periodic swimming}

The explicit expression in Eq. (2.7) for the asymptotic periodic swimming velocity $U_{sw}(t)$ can be put as
\begin{equation}
\label{3.1}U_{sw}(t)=\frac{1}{M(t)}\int^t_{-\infty}K(t,t')\mathcal{I}(t')\;dt',
\end{equation}
where the kernel $K(t,t')$ is positive, tends to zero as $t'\rightarrow-\infty$, increases monotonically as $t'$ increases, and has final value $K(t,t)=1$. We put
 \begin{equation}
\label{3.2}\gamma(t)=\frac{Z(t)}{M(t)},\qquad\gamma(t)=\overline{\gamma}+\delta\gamma(t),
\end{equation}
where $\delta\gamma(t)$ is periodic with vanishing mean value.
By putting $t'=t-\tau$ we can write the expression Eq. (2.7) in the alternative form
 \begin{equation}
\label{3.3}U_{sw}(t)=\frac{1}{M(t)H(t)}\int^\infty_0e^{-\overline{\gamma}\tau}H(t-\tau)\mathcal{I}(t-\tau)\;d\tau,
\end{equation}
with the periodic function
\begin{equation}
\label{3.4}H(t)=\exp{\bigg[\int^t_0\delta\gamma(t')\;dt'\bigg]}.
\end{equation}
We define the functions
 \begin{equation}
\label{3.5}V(t)=\frac{1}{M(t)H(t)},\qquad W(t)=H(t)\mathcal{I}(t).
\end{equation}
The mean swimming velocity is given by
 \begin{equation}
\label{3.6}\overline{U_{sw}}=\int^\infty_0e^{-\overline{\gamma}\tau}F(\tau)\;d\tau,
\end{equation}
with the function
 \begin{equation}
\label{3.7}F(\tau)=\overline{V(t)W(t-\tau)}.
\end{equation}
Both $V(t)$ and $W(t)$ are periodic with period $\mathcal{T}=2\pi/\omega$. Writing both functions as Fourier series of the form
 \begin{equation}
\label{3.8}V(t)=\sum^\infty_{n=-\infty} V_ne^{-in\omega t},\qquad V_n=\frac{1}{\mathcal{T}}\int^\mathcal{T}_0V(t)e^{in\omega t}\;dt,
\end{equation}
we have
 \begin{equation}
\label{3.9}F(\tau)=\sum^\infty_{n=-\infty} V^*_nW_ne^{in\omega\tau}.
\end{equation}
Hence the mean swimming velocity is given by
 \begin{equation}
\label{3.10}\overline{U_{sw}}=\frac{V_0W_0}{\overline{\gamma}}+2\;\mathrm{Re}\;\big[\sum^\infty_{n=1}\frac{V^*_nW_n}{\overline{\gamma}-in\omega}\big].
\end{equation}
In explicit calculations the sum converges rapidly. The expression can be generalized to the higher order Fourier coefficients of $U_{sw}(t)$ as
 \begin{equation}
\label{3.11}U_{sw,k}=\sum^\infty_{n=-\infty}\frac{V_{k-n}W_n}{\overline{\gamma}-in\omega}.
\end{equation}

We define relative coordinates $\{r_j\}$ with $j=1,...,N-1$ as
\begin{equation}
\label{3.12}r_1=x_2-x_1,\qquad r_2=x_3-x_2,...,\qquad r_{N-1}=x_N-x_{N-1}.
\end{equation}
The internal vibrations can be put as
\begin{equation}
\label{3.13}r_j(t)=d_{0j}+\varepsilon \xi_j(t),\qquad j=1,...,N-1,
\end{equation}
with equilibrium distances $\{d_{0j}\}$ and amplitude factor $\varepsilon$. The displacement vector $\du{d}(t)$ in Eq. (2.6) is defined by
\begin{equation}
\label{3.14}\du{d}=\varepsilon\du{T}^{-1}\cdot(0,\vc{\xi}),
\end{equation}
where $\du{T}$ is the matrix relating center and relative coordinates to the Cartesian coordinates, as given explicitly by Eqs. (2.1) and (3.12). The vector satisfies $\du{u}\cdot\du{d}=0$, as one sees by applying $\du{T}$ to Eq. (3.14).

In the asymptotic regime the individual sphere velocities, summarized in the $N$-vector
\begin{equation}
\label{3.15}\du{U}=U_{sw}\du{u}+\dot{\du{d}},
\end{equation}
are periodic functions of time. Hence the power, defined in Eq. (2.12), can be evaluated as the time average of the rate of dissipation over a period. Both the mean swimming velocity and the power are functions of the amplitude factor $\varepsilon$, and are of order $\varepsilon^2$ at small amplitude.

\section{\label{IV}Two-sphere swimmer}

As a first example we consider a two-sphere swimmer with spheres of radii $a$ and $b$ and mass densities $\rho_a,\rho_b$, centered at positions $x_1(t)$ and $x_2(t)$ on the $x$-axis. The relative position is $x(t)=x_2(t)-x_1(t)$. By convention $a>b$. A direct interaction ensures that $x(t)>a+b$. For $a=b$ and $\rho_a=\rho_b$ the swimming velocity vanishes by symmetry.

For two spheres the time-average of the impetus over a period vanishes. In periodic swimming the average of the first term in Eq. (2.5) vanishes obviously. The second term can be expressed as
\begin{equation}
\label{4.1}-\du{u}\cdot\vc{\zeta}\cdot\dot{\du{d}}=f(x)\;\frac{dx}{dt}.
\end{equation}
Its integral over a cycle is
\begin{equation}
\label{4.2}\oint f(x)\frac{dx}{dt}\;dt=\oint f(x)\;dx=0.
\end{equation}
Therefore for two spheres we have in periodic swimming
\begin{equation}
\label{4.3}\overline{D}=0,\qquad\overline{\mathcal{I}}=0.
\end{equation}
The average of each term on the right in Eq. (2.10) vanishes separately. In the Stokes limit the mean swimming velocity vanishes, because Eq. (2.8) takes the form
\begin{equation}
\label{4.4}U_{sw}(t)=\frac{f(x)}{Z(x)}\;\frac{dx}{dt},\qquad (\mathrm{Stokes}),
\end{equation}
so that the integral over a cycle vanishes as in Eq. (4.2).

In our explicit calculations we use approximate expressions for the hydrodynamic interactions. The friction matrix $\vc{\zeta}$ is found as the inverse of the mobility matrix $\vc{\mu}$, calculated in Oseen approximation \cite{11}. In Oseen approximation the friction matrix $\vc{\zeta}$ is given by
\begin{equation}
\label{4.5}\vc{\zeta}=\frac{12\pi\eta x}{4x^2-9ab}\left(\begin{array}{cc}2ax&-3ab\\-3ab&
2bx\end{array}\right).
\end{equation}
The mass matrix $\du{m}$ is found from potential flow theory in dipole approximation \cite{1},\cite{12}. In dipole approximation the mass matrix $\du{m}$ for the two-sphere swimmer is given by
\begin{equation}
\label{4.6}\du{m}=\frac{2\pi}{3x^6-3a^3b^3}\left(\begin{array}{cc}a^3[x^6(\rho+2\rho_a)+2a^3b^3(\rho-\rho_a)]&-3a^3b^3x^3\rho\\-3a^3b^3x^3\rho&
b^3[x^6(\rho+2\rho_b)+2a^3b^3(\rho-\rho_b)]\end{array}\right).
\end{equation}
The approximations are valid for distance $x$ between centers much larger than $a+b$.

We choose a time-dependent relative distance $x(t)$ given by
\begin{equation}
\label{4.7}x(t)=d+\varepsilon a \cos\omega t.
\end{equation}
In Fig. 1 we show the mean dimensionless swimming velocity $\overline{\hat{U}_{sw}}=\overline{U_{sw}}/(\omega a)$ for $b=a/2$ and $d=3a$, mass densities $\rho_a=\rho_b=\rho$ for $s=1$, as a function of the squared amplitude $\varepsilon^2$. The mean swimming velocity is negative, in the direction of the larger sphere. The variation with amplitude is nearly quadratic. The reduced mean swimming velocity is fitted by
\begin{equation}
\label{4.8}\frac{\overline{U_{sw}}}{\omega a}=0.00234\;\varepsilon^2-0.00006\;\varepsilon^4,
\end{equation}
in the range $0<\varepsilon<1$.
For a typical three-sphere swimmer the dependence is also nearly quadratic over a wide range \cite{3}. The streaming Reynolds number $Re_s$ used by Klotsa et al. \cite{6} is related to amplitude and scale number by
\begin{equation}
\label{4.9}Re_s=8\varepsilon^2s^2,
\end{equation}
so that for fixed scale number $s$ Fig. 1 may be regarded as a plot of swimming speed versus streaming Reynolds number.
In Fig. 2 we show the corresponding plot for the power. This shows also a nearly quadratic dependence on the amplitude. The reduced power is fitted by
\begin{equation}
\label{4.10}\frac{P}{\eta\omega^2a^3}=4.770\;\varepsilon^2+0.118\;\varepsilon^4,
\end{equation}
in the range $0<\varepsilon<1$. The Fourier coefficients of the swimming velocity decrease rapidly with increasing order. The first five absolute ratios for $\varepsilon=1$ are
\begin{equation}
\label{4.11}\{1,\bigg|\frac{U_{sw,1}}{\overline{U_{sw}}}\bigg|,\bigg|\frac{U_{sw,2}}{\overline{U_{sw}}}\bigg|,
\bigg|\frac{U_{sw,3}}{\overline{U_{sw}}}\bigg|,\bigg|\frac{U_{sw,4}}{\overline{U_{sw}}}\bigg|\}=\{1,66.81,3.37,0.60,0.12\}.
\end{equation}
Note that the ratio $|U_{sw,1}/\overline{U_{sw}}|$ is much larger than unity. The swimmer moves back and forth, making a little progress on average.

In Fig. 3 we plot the reduced mean dimensionless swimming velocity $\overline{U_{sw}}/(\omega a)$ for the above example with $\varepsilon=1$ as a function of the square scale number $s^2$. In the Stokes limit the mean swimming velocity vanishes, as found below Eq. (4.4). For large $s$ the function decays in proportion to $1/s^2$.

We define the dimensionless efficiency as
\begin{equation}
\label{4.12}E_T=\eta\omega a^2\frac{|\overline{U_{sw}}|}{P}.
\end{equation}
In Fig. 4 we show the efficiency for the above example with $\varepsilon=1$ as a function of scale number $s$. The efficiency is maximal at $s=1.230$ and there takes the value $0.000505$.

In their experiments and computer simulations for scale number $s\approx 40$ Klotsa et al. \cite{2} find a sharp increase in mean swimming velocity at streaming Reynolds number $Re_s\approx 20$  corresponding to $\varepsilon\approx 0.05$. This corresponds to a regime far to the right in Figs. 3 and 4. For larger values of the amplitude the mean swimming velocity appears to depend only on $Re_s$, but to be independent of frequency for the same $Re_s$. The behavior is clearly not covered by our theory, and suggests that a new mechanism sets in with a transcritical bifurcation type instability \cite{7}, as in Haken's model of the laser.

\section{\label{V}Three-sphere chain}

As a second example we consider a chain of three spheres of equal radius $a$ and mass density $\rho_a$. The relative positions of centers are $r_1(t)=x_2(t)-x_1(t)$ and $r_2(t)=x_3(t)-x_2(t)$. We discussed this model with harmonic elastic interactions earlier in the Stokes limit \cite{4},\cite{9} and with inertia \cite{1} on the basis of Eq. (2.3). Here we take a kinematic point of view and assume that the relative positions vary harmonically. In analogy to Eq. (4.7) we consider relative positions
\begin{equation}
\label{5.1}r_1(t)=d+\varepsilon a \cos\omega t,\qquad r_2(t)=d+\varepsilon a\xi \cos(\omega t-\varphi),
\end{equation}
with equilibrium distance $d$, relative magnitude $\xi$, and phase difference $\varphi$. This can be used directly in Eq. (2.4). Hence we evaluate the asymptotic periodic swimming velocity $U_{sw}(t)$, as given by Eq. (2.7), and the rate of dissipation $\mathcal{D}(t)$, as given by Eq. (2.11). This three-sphere swimmer is much more efficient than the two-sphere swimmer considered above. We shall study the efficiency, as defined in Eq. (4.12), as a function of the scale number $s$.

The stroke is specified by the amplitude factor $\varepsilon$, the relative magnitude $\xi$, and the phase difference $\varphi$. We choose the latter two values such that the efficiency is optimized in the Stokes limit in the bilinear theory. To second order in $\varepsilon$ the optimum stroke is found from an eigenvalue problem, which we solved earlier in analytic form \cite{4}. We found that in the bilinear theory applied to a fluid  with inertia and neutrally buoyant spheres the optimum stroke is nearly the same \cite{1}.

In the following we consider neutrally buoyant spheres with $\rho_a=\rho$ and $d=3a$. The friction matrix is evaluated in Oseen approximation and the mass matrix is evaluated in dipole approximation, as for the two-sphere system. We examine first to what extent the bilinear theory is valid. In Fig. 5 we show the reduced mean swimming velocity $\overline{U_{sw}}/(\omega a)$ as a function of $\varepsilon^2$ for $s=1$. The dependence is nearly quadratic and the curve is fitted by
\begin{equation}
\label{5.2}\frac{\overline{U_{sw}}}{\omega a}=0.044\;\varepsilon^2+0.008\;\varepsilon^4,
\end{equation}
in the range $0<\varepsilon<1$. In Fig. 6 we show the corresponding plot for the power. This also shows a nearly quadratic dependence on amplitude. The reduced power is fitted by
\begin{equation}
\label{5.3}\frac{P}{\eta\omega^2a^3}=18.68\;\varepsilon^2+1.76\;\varepsilon^4,
\end{equation}
in the range $0<\varepsilon<1$. The first five absolute ratios of the Fourier coefficients for $\varepsilon=1,\;s=1$ are
\begin{equation}
\label{5.4}\{1,\bigg|\frac{U_{sw,1}}{\overline{U_{sw}}}\bigg|,\bigg|\frac{U_{sw,2}}{\overline{U_{sw}}}\bigg|,
\bigg|\frac{U_{sw,3}}{\overline{U_{sw}}}\bigg|,\bigg|\frac{U_{sw,4}}{\overline{U_{sw}}}\bigg|\}=\{1,0.868,0.073,0.037,0.009\}.
\end{equation}
Per beat the three-sphere chain makes much more progress than the two-sphere system.

In Fig. 7 we show the reduced mean swimming velocity $\overline{U_{sw}}/(\omega a)$ as a function of $s^2$ for $\varepsilon=1$. In Fig. 8 we show the efficiency $E_T$ as a function of $s^2$ for $\varepsilon=1$. The efficiency varies little with $s$ in the whole range studied, as we found earlier for the maximum eigenvalue in the bilinear theory \cite{1}. The three-sphere chain is more than ten times more efficient than the comparable two-sphere system given by the maximum in Fig. 4.

In the Stokes limit we find $\overline{U_{sw}}/(\omega a)=0.0552$ and $E_T=0.0027$, again at $\varepsilon=1$. For the maximum eigenvalue in the bilinear theory we find in the Stokes limit $\lambda_+=0.0025$, as given by Eq. (6.5) in Ref. 4.

Finally, it is of interest to compare the kinetic energy of the swimmer, including added mass, with the total kinetic energy in the asymptotic periodic regime. Thus we define
\begin{equation}
\label{5.5}\mathcal{K}_{sw}(t)=\frac{1}{2}\;M(t)U_{sw}(t)^2,\qquad\mathcal{K}(t)=\frac{1}{2}\;\du{U}\cdot\du{m}\cdot\du{U}.
\end{equation}
In Fig. 9 we show the ratio of the two quantities as a function of time for the three-sphere chain studied above for amplitude $\varepsilon=1$ and squared scale number $s^2=10$. The figure shows that the kinetic energy $\mathcal{K}_{sw}(t)$ is small compared to the total $\mathcal{K}(t)$ over the whole period. For comparison we show also the swimming velocity $U_{sw}(t)$ and the total kinetic energy $\mathcal{K}(t)$ in suitably chosen units. This shows that the kinetic energy varies more strongly than the swimming velocity.

\section{\label{V}Discussion}

The analysis shows that the swimming velocity is the result of a remarkable interplay of the effects of friction and added mass. For the discrete particle model with given applied forces the motions of the separate spheres can be analyzed in detail. In our calculations we concentrated on the asymptotic periodic swimming velocity. This can be studied from a kinematic point of view with specified time-dependence of the relative distances of the spheres, as in Eqs. (4.7) and (5.1). In these calculations it is not necessary to know the applied forces and the direct interaction potential. The required power can be studied in the same context.

In our model calculation for the two-sphere system we find a dependence of the mean swimming velocity on the amplitude of excitation, but no sharp transition as in the experiments of Klotsa et al. \cite{6}.  Presumably with more accurate hydrodynamic interactions the dependence on amplitude becomes more pronounced. In the lubrication region the hydrodynamic interactions depend greatly on distance. This suggests that for amplitude $\varepsilon\approx(d-a-b)/a$ in Eq. (4.7) there is a strong increase in the mean swimming velocity. However, the above estimate suggests that the transition seen in experiment occurs also at small amplitude. The experiments indicate that a new mechanism sets in at $Re_s\approx 20$, where $Re_s$ is the streaming Reynolds number defined in Eqs. (2.9) and (4.9).

Transient effects can be studied on the basis of the postulated equations of motion Eq. (2.2) with specified direct interaction, for example a harmonic spring interaction. One can associate a flow pattern with the instantaneous positions and velocities, given by a superposition of Oseen flow patterns corresponding to the forces on the spheres and potential flow patterns corresponding to the sphere velocities. More accurate hydrodynamic interactions lead to more complicated flow patterns, with the effect of higher order multipoles included. It would be of interest to compare the motion of the spheres and the corresponding flow patterns with computer simulations.

\newpage

\section*{Figure captions}

\subsection*{Fig. 1}
Plot of the dimensionless mean swimming velocity $\overline{U_{sw}}/(\omega a)$ of the two-sphere swimmer as a function of the square amplitude $\varepsilon^2$ for $\rho_a=\rho_b=\rho$, $b=a/2$ and $d=3a$ at scale number $s=1$.

\subsection*{Fig. 2}
Plot of the dimensionless mean power $P/(\varepsilon^2\eta\omega^2 a^3)$  as a function of the square amplitude $\varepsilon^2$ for the same model as in Fig. 1.

\subsection*{Fig. 3}
Plot of the dimensionless mean swimming velocity $\overline{U_{sw}}/(\omega a)$ of the two-sphere swimmer as a function of the square scale number $s^2$ for $\rho_a=\rho_b=\rho$, $b=a/2$ and $d=3a$ at amplitude factor $\varepsilon=1$.

\subsection*{Fig. 4}
Plot of the efficiency $E_T$ of the two-sphere swimmer as a function of the square scale number $s^2$ for $\rho_a=\rho_b=\rho$, $b=a/2$ and $d=3a$ at amplitude factor $\varepsilon=1$.

\subsection*{Fig. 5}
Plot of the dimensionless mean swimming velocity $\overline{U_{sw}}/(\omega a)$ of the three-sphere chain as a function of the square amplitude $\varepsilon^2$ for $\rho_a=\rho_b=\rho$ and $d=3a$ at scale number $s=1$.

\subsection*{Fig. 6}
Plot of the dimensionless mean power $P/(\varepsilon^2\eta\omega^2 a^3)$  as a function of the square amplitude $\varepsilon^2$ for the same model as in Fig. 5.

\subsection*{Fig. 7}
Plot of the dimensionless mean swimming velocity $\overline{U_{sw}}/(\omega a)$ of the three-sphere chain as a function of the square scale number $s^2$ for $\rho_a=\rho_b=\rho$ and $d=3a$ at amplitude factor $\varepsilon=1$.

\subsection*{Fig. 8}
Plot of the efficiency $E_T$ of the three-sphere chain as a function of the square scale number $s^2$ for $\rho_a=\rho_b=\rho$ and $d=3a$ at amplitude factor $\varepsilon=1$.

\subsection*{Fig. 9}
Plot of the ratio of the total kinetic energy $\mathcal{K}(t)$ and the kinetic energy of the swimmer $\mathcal{K}_{sw}(t)$, as defined in Eq. (5.5), as a function of time for the three-sphere swimmer at amplitude factor $\varepsilon=1$ and squared scale number $s^2=10$ (drawn curve). We compare with the total kinetic energy $\mathcal{K}(t)$ (short dashes) and the swimming velocity $U_{sw}(t)$ (long dashes) in suitably chosen units.

\clearpage
\newpage
\setlength{\unitlength}{1cm}
\begin{figure}
 \includegraphics{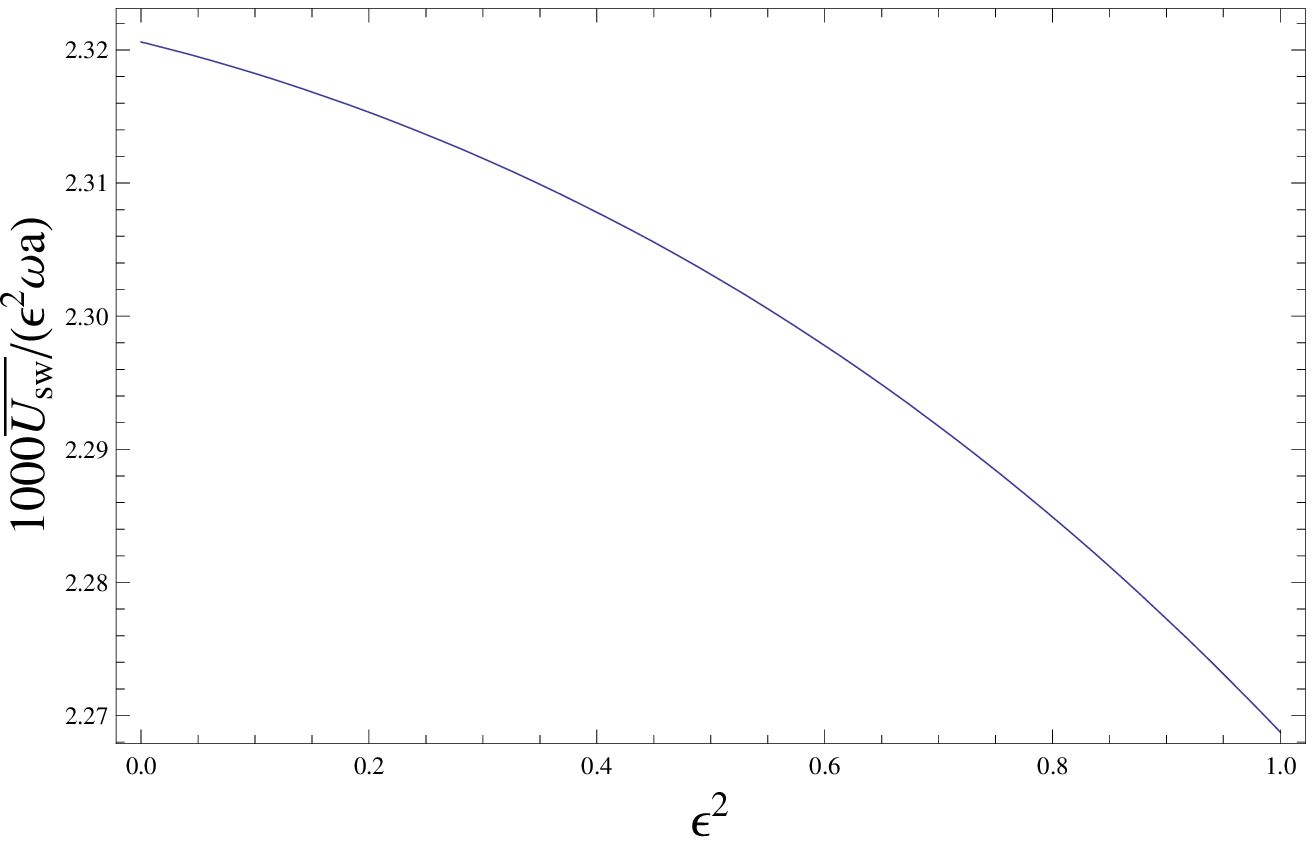}
   \put(-9.1,3.1){}
\put(-1.2,-.2){}
  \caption{}
\end{figure}
\newpage
\clearpage
\newpage
\setlength{\unitlength}{1cm}
\begin{figure}
 \includegraphics{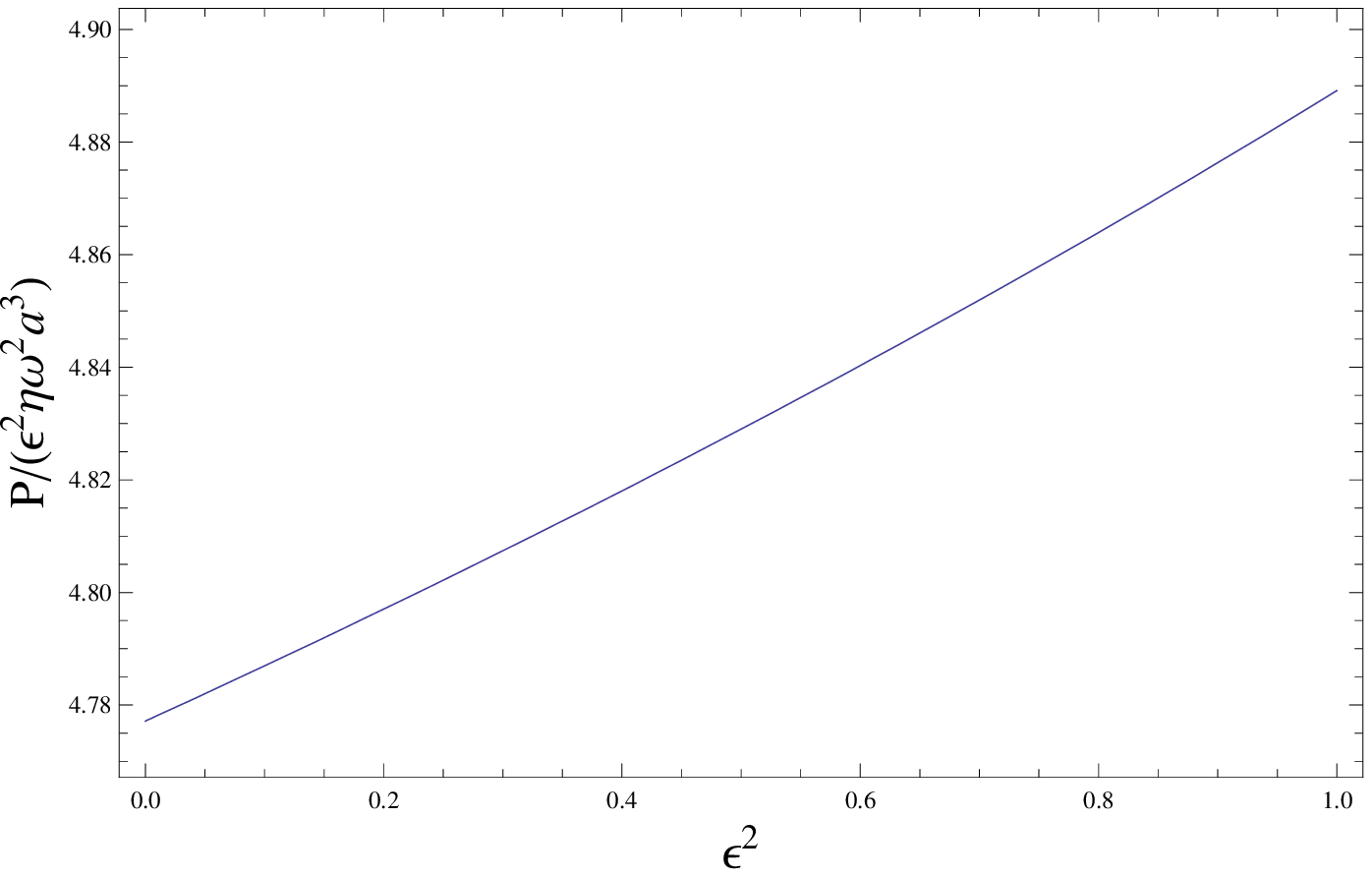}
   \put(-9.1,3.1){}
\put(-1.2,-.2){}
  \caption{}
\end{figure}
\newpage
\clearpage
\newpage
\setlength{\unitlength}{1cm}
\begin{figure}
 \includegraphics{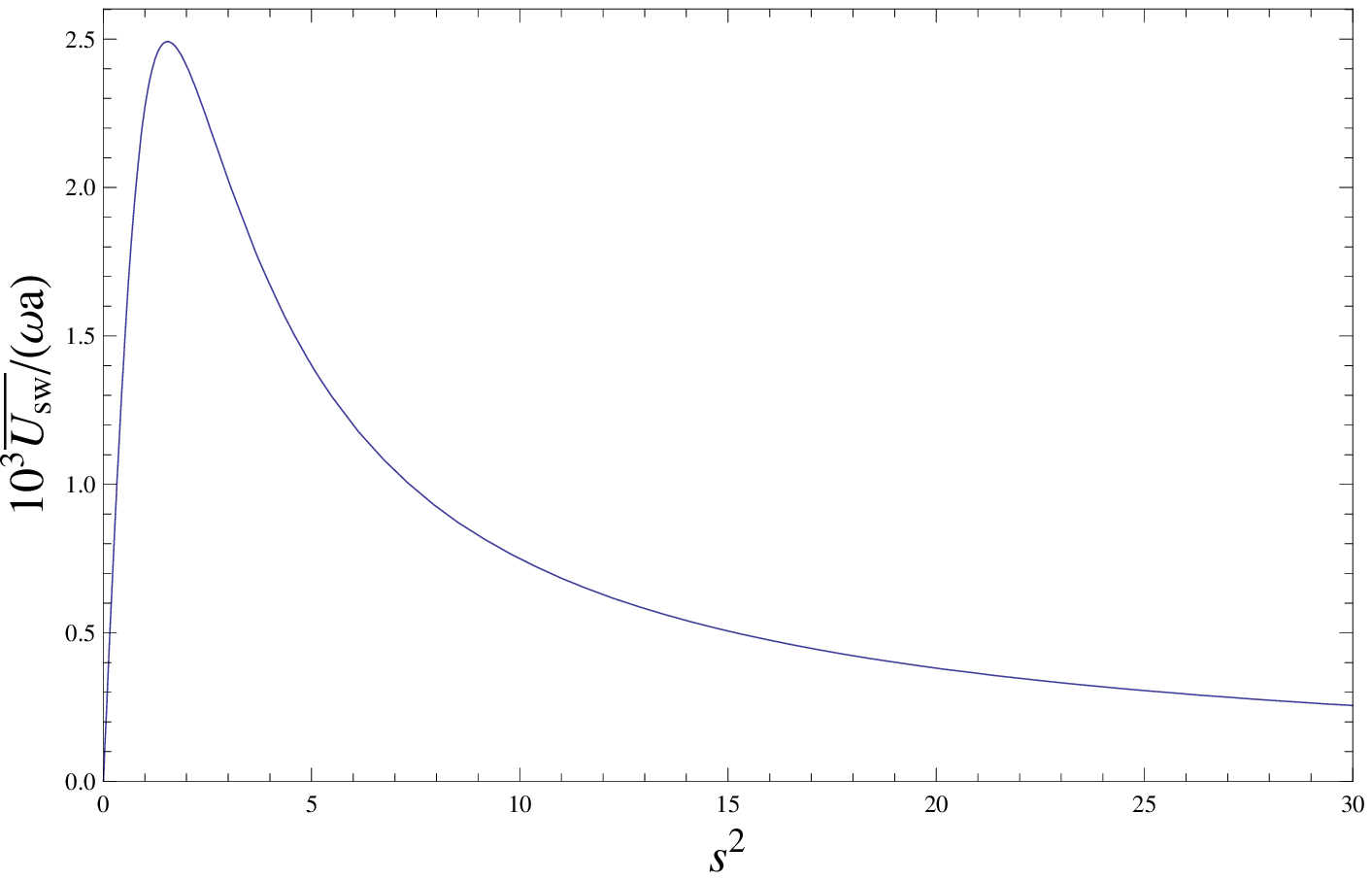}
   \put(-9.1,3.1){}
\put(-1.2,-.2){}
  \caption{}
\end{figure}
\newpage
\clearpage
\newpage
\setlength{\unitlength}{1cm}
\begin{figure}
 \includegraphics{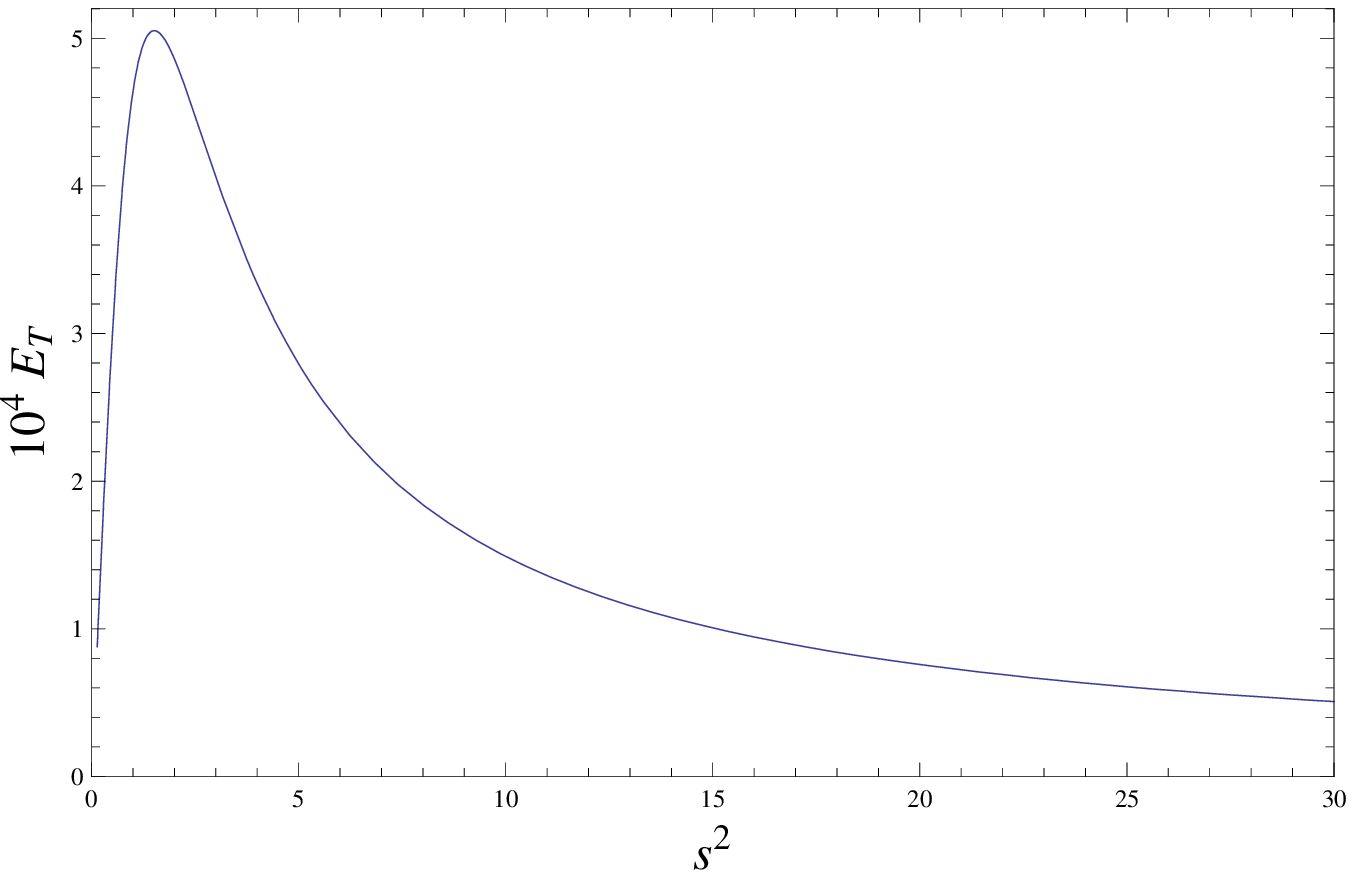}
   \put(-9.1,3.1){}
\put(-1.2,-.2){}
  \caption{}
\end{figure}
\newpage
\clearpage
\newpage
\setlength{\unitlength}{1cm}
\begin{figure}
 \includegraphics{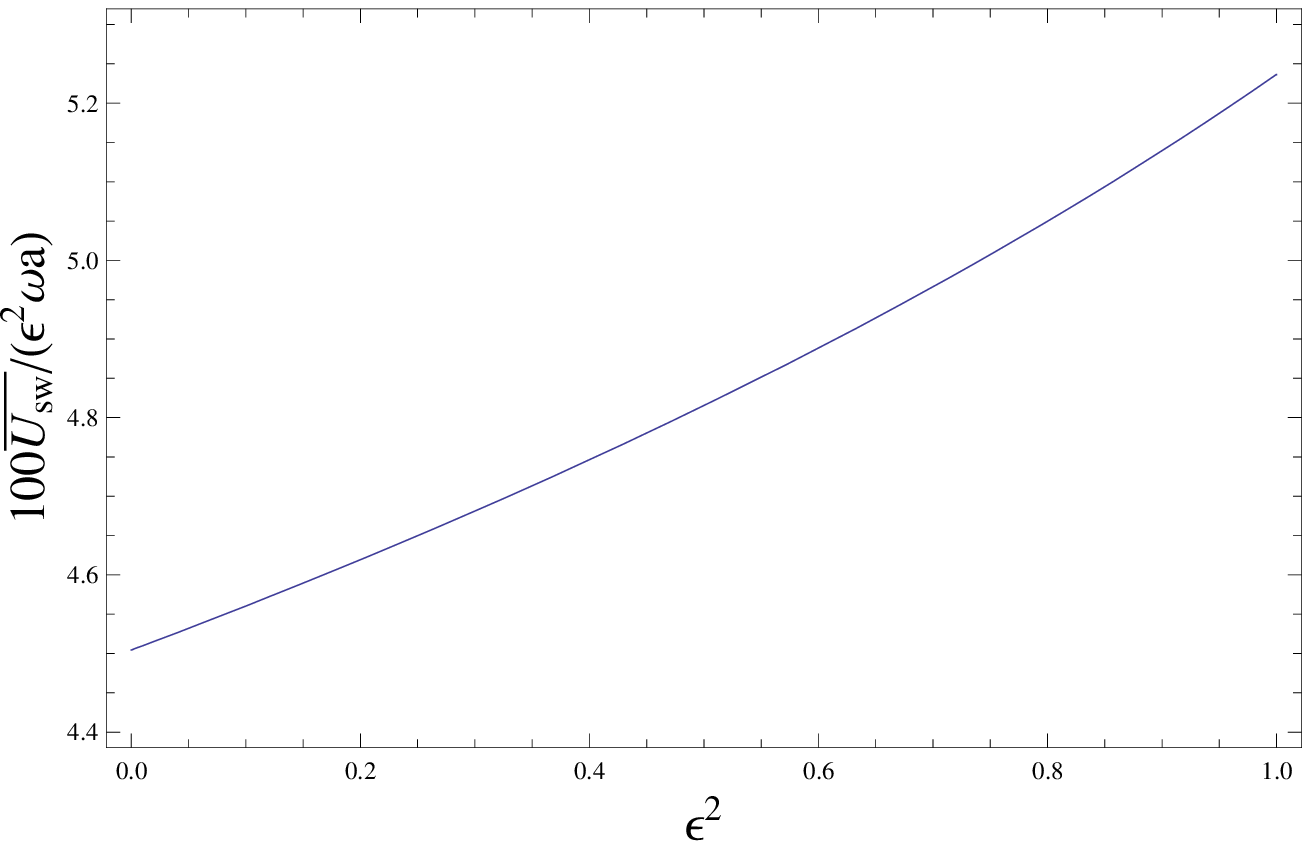}
   \put(-9.1,3.1){}
\put(-1.2,-.2){}
  \caption{}
\end{figure}
\newpage
\clearpage
\newpage
\setlength{\unitlength}{1cm}
\begin{figure}
 \includegraphics{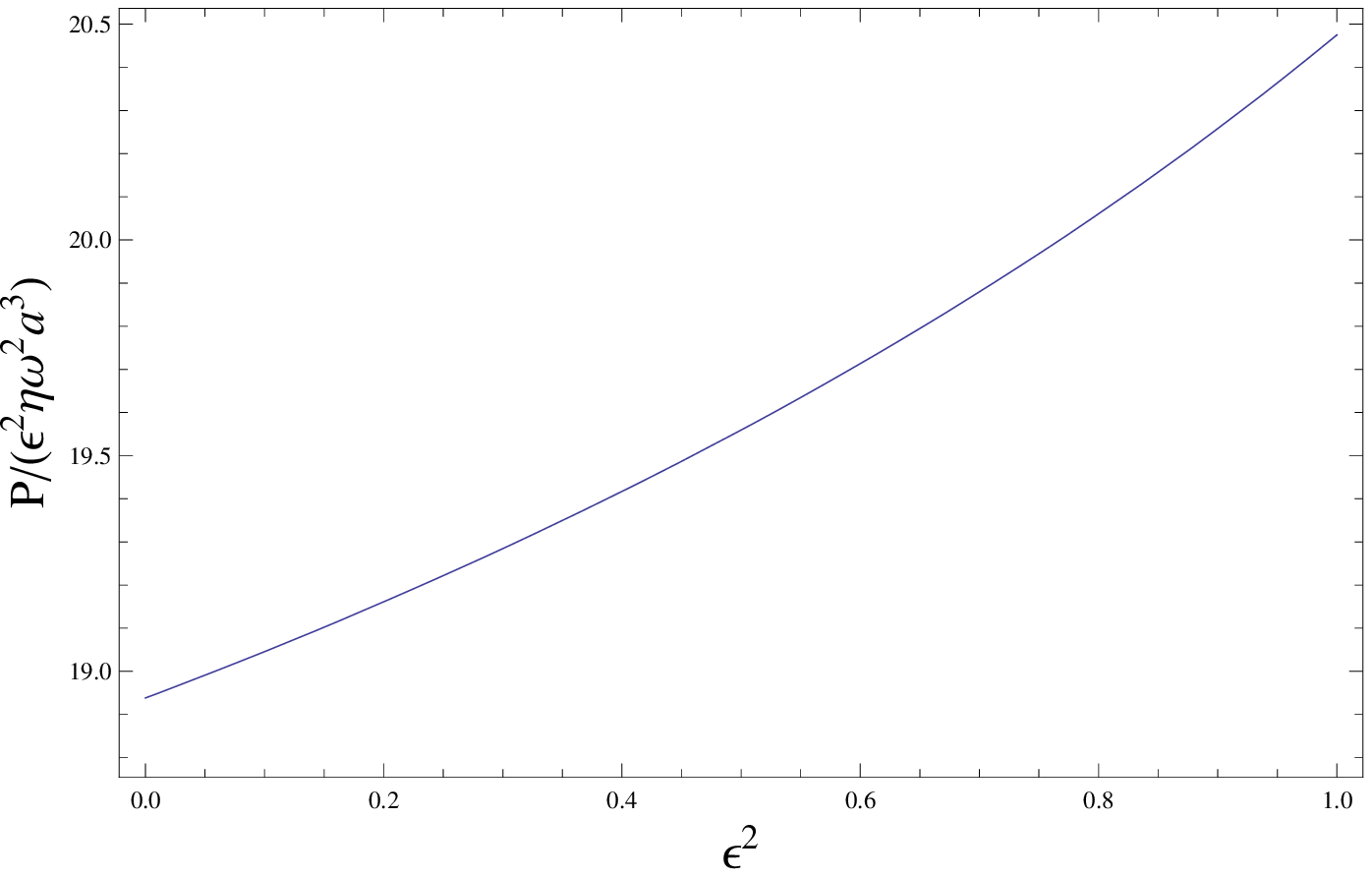}
   \put(-9.1,3.1){}
\put(-1.2,-.2){}
  \caption{}
\end{figure}
\newpage
\clearpage
\newpage
\setlength{\unitlength}{1cm}
\begin{figure}
 \includegraphics{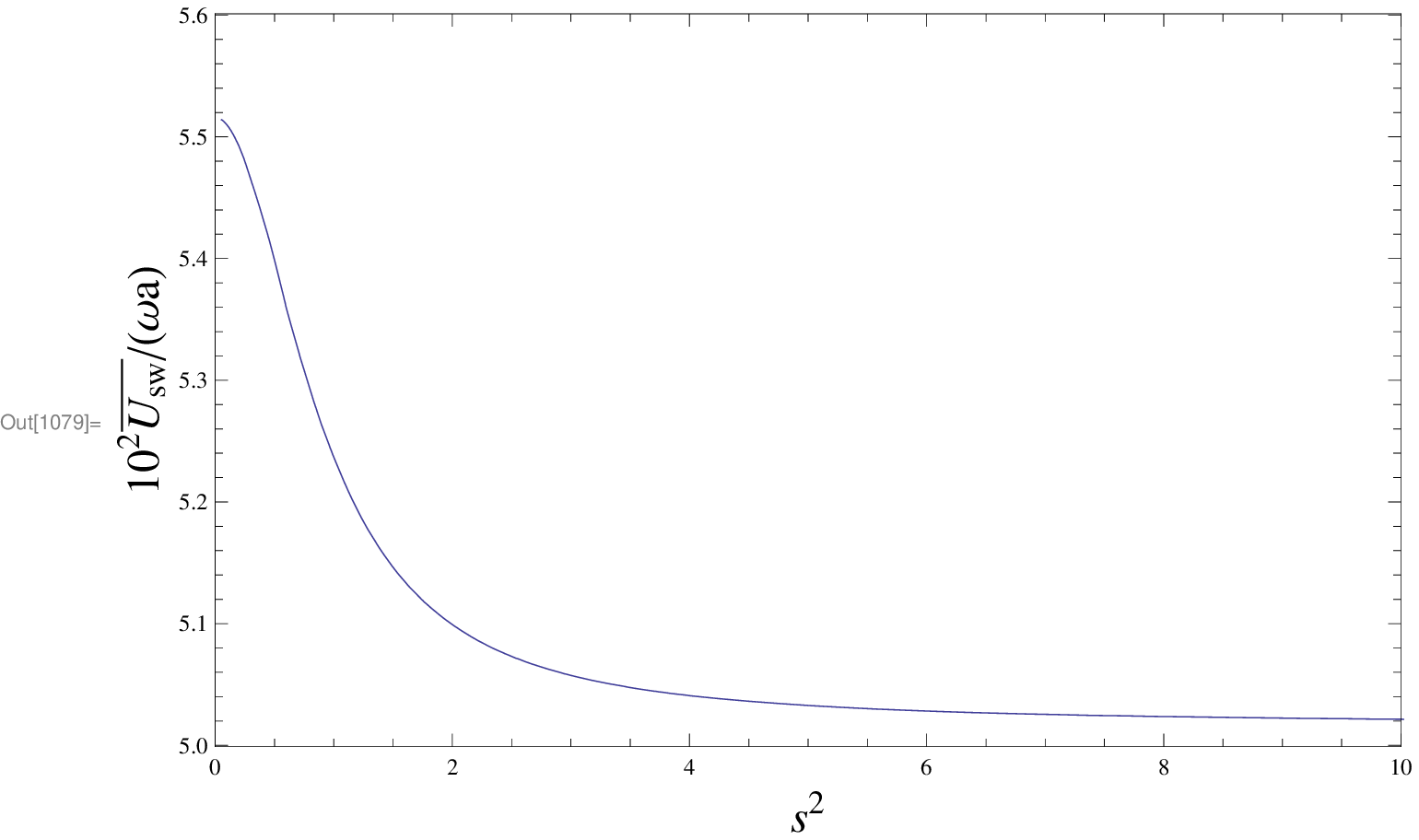}
   \put(-9.1,3.1){}
\put(-1.2,-.2){}
  \caption{}
\end{figure}
\newpage
\clearpage
\newpage
\setlength{\unitlength}{1cm}
\begin{figure}
 \includegraphics{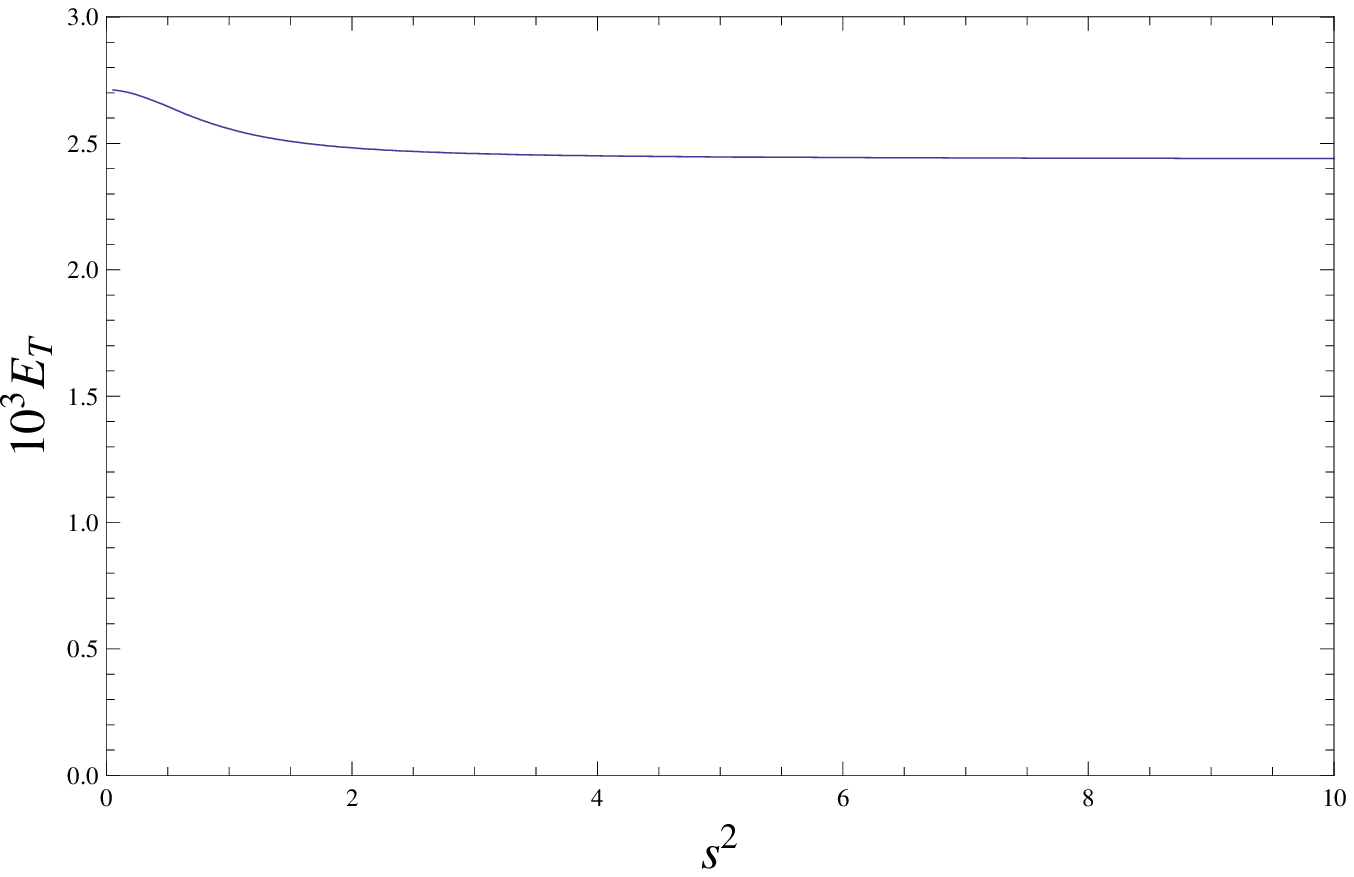}
   \put(-9.1,3.1){}
\put(-1.2,-.2){}
  \caption{}
\end{figure}
\newpage
\clearpage
\newpage
\setlength{\unitlength}{1cm}
\begin{figure}
 \includegraphics{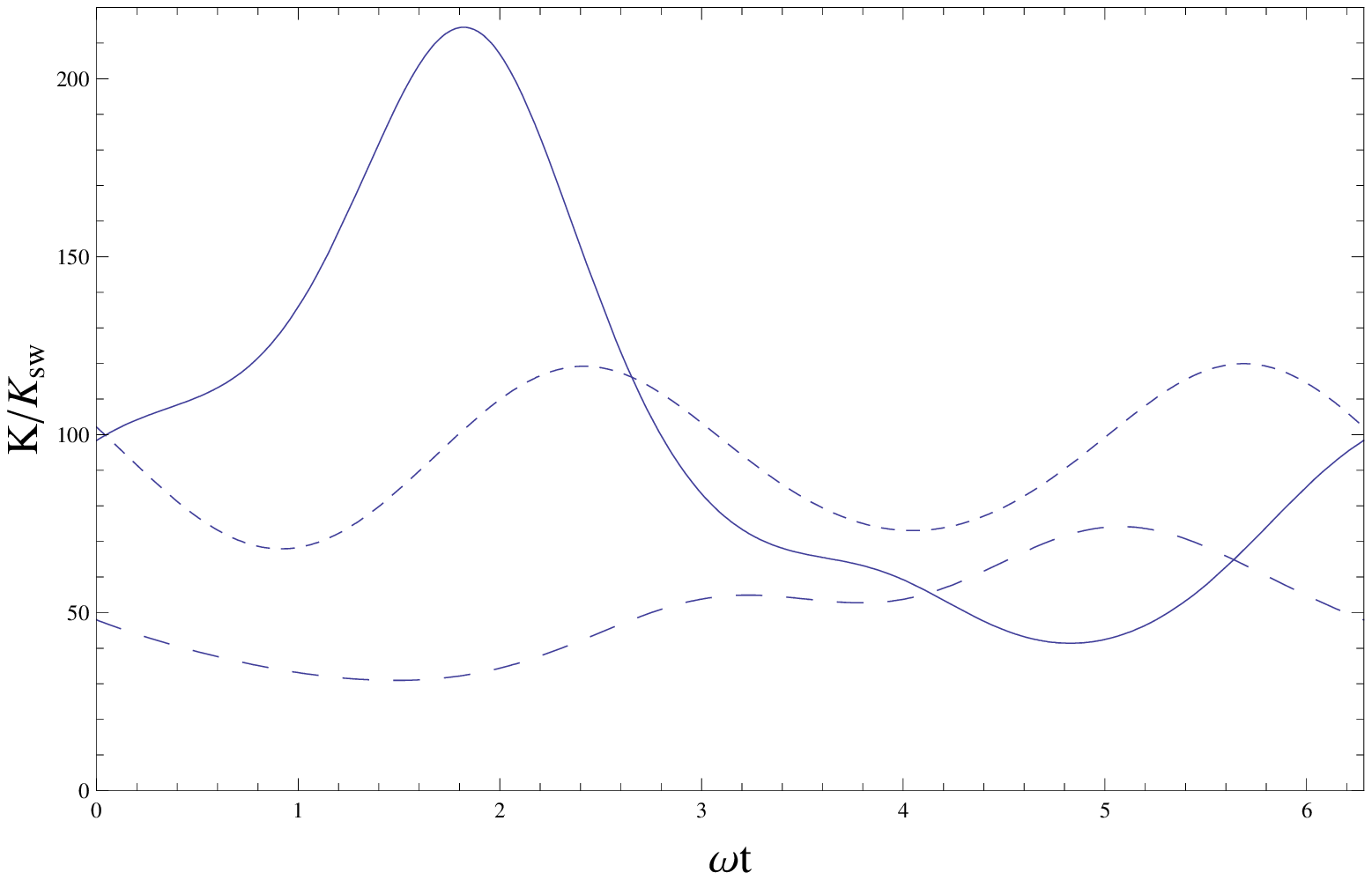}
   \put(-9.1,3.1){}
\put(-1.2,-.2){}
  \caption{}
\end{figure}
\end{document}